\documentclass[a4paper,10pt]{article}
\usepackage[utf8]{inputenc}
\usepackage{amsmath,amsthm,amssymb}
\usepackage{enumerate}
\usepackage{hyperref}
\usepackage{doi}
\usepackage{graphicx}
\usepackage{subfigure}
\usepackage{xcolor}
\usepackage{authblk} \usepackage{cite}
\usepackage{dsfont} \usepackage{stmaryrd} \usepackage{bbm}
\usepackage[a4paper,left=3cm,right=3cm,top=3cm,bottom=3cm]{geometry}

\theoremstyle{definition}

\theoremstyle{definition}
\newtheorem{rmk}{Remark}[]

\newcommand{\RR}{\mathbb{R}}  \DeclareMathOperator{\sgn}{sgn} \DeclareMathOperator{\hs}{\mathcal{N}}\DeclareMathOperator{\dref}{\mathcal{D}_{ref}} 

\newcommand{\inproduct}[2]{\left\langle #1 , #2 \right\rangle}
\newcommand{\jump}[1]{\left\llbracket #1 \right\rrbracket}
\newcommand{\nsigma}{n_\Sigma}
\newcommand{\ddt}[1]{\frac{d #1}{dt}}

\newcommand{\transpose}{\mathsf{T}} \newcommand{\ndomega}{n_{\partial\Omega}}

\newcommand{\clspeed}{V_\Gamma}
\newcommand{\sigmawet}{\sigma_{\text{w}}}
\newcommand{\Ca}{\text{Ca}} \newcommand{\normalspeed}{V_\Sigma}
 \newcommand{\change}[1]{#1} \newcommand{\todo}[1]{} 

\title{An analytical study of capillary rise dynamics:\\ Critical conditions and hidden oscillations}
\author[1]{M.~Fricke\thanks{Corresponding Author: \url{fricke@mma.tu-darmstadt.de}, Technische Universität Darmstadt, Peter-Grünberg-Str.~10, 64287 Darmstadt, Germany.}}
\author[1]{E.~Ouro-Koura}
\author[1]{S.~Raju}
\author[2]{R.~von~Klitzing}
\author[3]{J.~De~Coninck}
\author[1]{D.~Bothe}

\affil[1]{Mathematical Modeling and Analysis Group, TU Darmstadt, Germany}
\affil[2]{Soft Matter at Interfaces, TU Darmstadt, Germany}
\affil[3]{Transfers, Interfaces and Processes, Université libre de Bruxelles, Belgium}

\date{} 

\usepackage{xr}

\begin{document}

\maketitle

\begin{abstract}
The rise of a liquid column inside a thin capillary against the action of gravity is a prototypical example of a dynamic wetting process and plays an important role for applications but also for fundamental research in the area of multiphase fluid dynamics. Since the pioneering work by Lucas and Washburn, many research articles have been published which aim at a simplified description of the capillary rise dynamics using complexity-reduced models formulated as ordinary differential equations. Despite the fact that these models are based on profound simplifications, they may still be able to describe the essential physical mechanisms and their interplay. In this study, we focus on the phenomenon of oscillations of the liquid column. The latter has been observed experimentally for liquids with sufficiently small viscosity leading to comparably small viscous dissipation. Back in 1999, Qu{\'e}r{\'e} et al.\ formulated a condition for the appearance of rise height oscillations for an ODE model introduced by Bosanquet in 1923. This model has later been extended to include further dissipative mechanisms. In this work, we extend the mathematical analysis to a larger class of models including additional channels of dissipation. We show that Qu{\'e}r{\'e}'s critical condition is generalized to $\Omega + \beta < 2$, where $\Omega$ was introduced earlier and $\beta$ is an additional non-dimensional parameter describing, e.g., contact line friction. A quantitative prediction of the oscillation dynamics is achieved from a linearization of the governing equations. We apply the theory to experimental data by Qu{\'e}r{\'e} et al.\ and, in particular, reveal the oscillatory behavior of dynamics for the nearly critically damped case of ethanol.
 \end{abstract}

\newline
This preprint was submitted and accepted for publication in \emph{Physica D: Nonlinear Phenomena}.\newline When citing this work, please refer to the journal article: \textbf{DOI:} \href{https://www.doi.org/10.1016/j.physd.2023.133895}{10.1016/j.physd.2023.133895}.

\section{Introduction}
\begin{figure}[ht] 
\centering
\includegraphics[width=0.25\columnwidth]{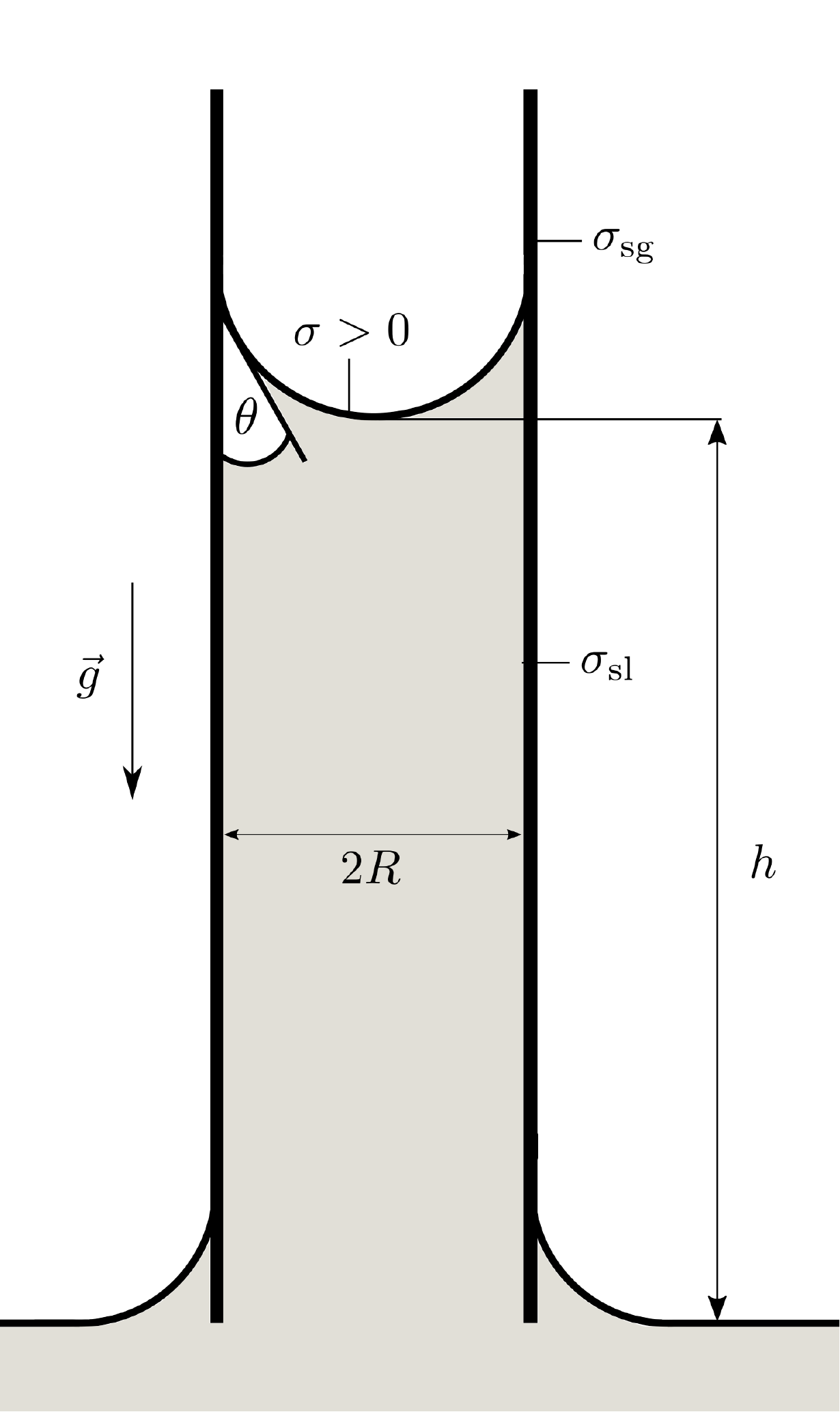}
\caption{The classical capillary rise problem.}\label{fig:capillaryRise-setup}
\end{figure}

The problem of penetration of liquid into a pore or a thin capillary has attracted the attention of researchers for more than one hundred years (see the pioneering work by Lucas \cite{Lucas1918}, Washburn \cite{Washburn1921} and Bosanquet \cite{Bosanquet1923}). It is important for applications such as oil recovery or flow in porous media, but also for the development of mathematical theories of dynamic wetting. From the theoretical point of view, the problem is particularly appealing because of its geometrical simplicity. Neglecting the curvature of the free surface, it is straightforward to derive an expression for the stationary height (measured with respect to the, ideally infinite, liquid reservoir) of a meniscus rising against gravity in a cylindrical capillary  (see Fig.~\ref{fig:capillaryRise-setup}):
\begin{align}\label{eqn:jurins_height}
h_0 = \frac{2\sigma \cos \theta_0}{\rho g R}.
\end{align}
Here $\sigma$ denotes the surface tension of the liquid-gas interface, $\theta_0$ is the equilibrium contact angle of the liquid in contact with the capillary walls, $\rho$ is the density of the liquid, $g$ is the gravitational acceleration and $R$ is the radius of the capillary tube. Equation \eqref{eqn:jurins_height} is known as ``Jurin's height'' named after James Jurin who discovered it back in the year 1718. If necessary, the formula can be improved by adding a correction term which takes into account the interface curvature (see \cite{Gruending2019,Gruending2020b}).\\
\\
Despite the simplicity of the setup, it is still very challenging to (analytically or numerically) predict the \emph{dynamics} of the process. This difficulty is essentially caused by the multiscale nature of the dynamic wetting process. It has been shown that, in the classical sharp-interface two-phase flow setting, the no-slip boundary condition cannot describe a moving contact line \cite{Huh1971}.  In fact, the no-slip condition would imply an infinite dissipation rate, a finding known as the ``Huh-Scriven paradox''. Consequently, different boundary conditions are necessary for a continuum mechanical description (see Section~\ref{sec:continuum-theory} for more details).  

\paragraph{Complexity-reduced models for the capillary rise process:} In the seminal paper from 1923, Bosanquet \cite{Bosanquet1923} introduced the following model for the dynamics of the rise height $h=h(t)$ with respect to the liquid reservoir
\begin{equation}\label{eqn:bosanquet_model}
2 \pi R \sigma \cos \theta_0 =  8 \pi \eta h \dot{h} + \pi R^2 \rho \ddt{} (h \dot{h}) + \pi R^2 \rho g h.
\end{equation}
It is derived as a balance of forces using significant simplifications regarding the geometry and flow structure. It is assumed that the liquid column can be approximated by a cylinder with radius $R$ and height $h$ moving with an average velocity $\dot{h}$, one obtains formulas for the mass $M$, the momentum $P$ and the gravitational force $f_g$ acting on the column
\[ M(h) = \pi R^2 h \rho, \quad P(h,\dot{h}) = M(h) \dot{h}, \quad F_g = - M(h) g = - \pi R^2 h \rho g.  \]
The driving force of the process is related to the gain in surface energy from the wetting of the solid boundary. We use the symbol
\[ \sigmawet = \sigma_{\text{sl}} - \sigma_{\text{sg}}  \]
to denote the specific energy per unit area for wetting, which is nothing but the difference of the surface energy (surface tensions) of the solid-liquid and solid-gas interface, respectively. In the following, we assume the solid to be hydrophilic, i.e.\ $\sigmawet < 0$ which leads to a positive stationary rise height. With this notation, we may model the driving force of the process as 
\[ F_{\text{w}} = -\sigmawet \frac{\text{d}A}{\text{d}h} = - \sigmawet 2 \pi R. \]
In the case of partial wetting, we may rewrite $F_{\text{w}}$ using the Young equation \cite{Young.1805} for the equilibrium contact angle $\theta_0$, i.e.\
\[ \sigma \cos \theta_0 + \sigmawet = 0, \]
leading to $F_{\text{w}} = 2 \pi R \sigma \cos \theta_0$. Moreover, a viscous resistance force can be computed assuming a Hagen–Poiseuille flow with average velocity $\dot{h}$ throughout the entire column. The resulting force is
\[ F_\eta = - 8 \pi \eta h \dot{h}, \]
where $\eta$ denotes the dynamic viscosity. Now, equation \eqref{eqn:bosanquet_model} is obtained from balancing the forces according to
\[ \ddt{} P(h,\dot{h}) =  F_{\text{w}} + F_\eta + F_g.  \]

\paragraph{The analysis by Qu{\'e}r{\'e} et al.:} To proceed, it is useful to rewrite \eqref{eqn:bosanquet_model} in dimensionless form. Following \cite{Quere1997,Quere1999} we choose the dimensionless variables
\begin{align}
H = h/h_0 \quad \text{and} \quad s = t/\tau \quad \text{where}  \quad \tau = \sqrt{\frac{h_0}{g}}.
\end{align}
The quantity $h_0$ denotes the unique stationary solution of \eqref{eqn:bosanquet_model}, i.e. the Jurin's height \eqref{eqn:jurins_height}. Rewriting \eqref{eqn:bosanquet_model} in  non-dimensional form leads to
\begin{align}\label{eqn:bosanquet_model_dimless}
(H H')' + \Omega \, H H' + H - 1 = 0.
\end{align}
We observe that there is one non-dimensional parameter
\begin{equation}\label{eqn:definition_omega}
\Omega:= \sqrt{\frac{128 \eta^{2} \sigma \cos \theta_0}{R^{5} \rho^{3} g^{2}}}=8 \sqrt{2 \cos\theta_0} \, \frac{\mathrm{Oh}}{\mathrm{Bo}} 
\end{equation}
governing behavior of solutions of \eqref{eqn:bosanquet_model}. Here the Ohnesorge number $\text{Oh}$ and the Bond number $\text{Bo}$ are defined as
\[ \text{Oh} = \frac{\eta}{\sqrt{R \rho \sigma}} \quad \text{and} \quad \text{Bo} = \frac{\rho g R^2}{\sigma}, \quad \text{respectively.} \]
Using a linearization close to stationary state based on the substitution
\begin{align}\label{eqn:simple_linearization}
H(s) = 1 + \varepsilon(s) 
\end{align}
for $|\varepsilon| \ll 1$, Qu{\'e}r{\'e} et al.\ showed \cite{Quere1999} that there is a regime transition from oscillatory to monotonic rise at the critical parameter
\[ \Omega_c = 2. \]
Hence, an oscillatory solution is expected for $\Omega < 2$, i.e.\ for a liquid with sufficiently low viscosity.\newline
\newline
The original mathematical analysis by Qu{\'e}r{\'e} et al.\ was later extended with more details regarding the physical regimes and mathematical properties in \cite{Zhmud2000,Lorenceau2002,Fries2008a,Fries2008,Das2013,Marston2018,Plociniczak2018,Zhang2018},\change{\cite{Lunowa2022}}. P{\l}ociniczak and {\'{S}}wita{\l}a gave a rigorous proof of the critical condition and the well-posedness of the model. Using a linearization of the equations based on \eqref{eqn:simple_linearization}, Marston et al.\ derived the oscillation period for the classical model and demonstrated an analogy to a damped harmonic oscillator.\newline
Different physical mechanisms, which are not included in Bosanquet's model, have been shown to be relevant for the capillary rise dynamics. In particular, the dissipation caused by entrance or exit of the fluid at the inlet/outlet of the pipe is closely investigated in \cite{Lorenceau2002,Ramakrishnan2019,Wang2019,Plociniczak2021}. These effects can even be the dominant ones in the system if the viscosity is very low \cite{Lorenceau2002}. \change{Fiorini et al.\ \cite{Fiorini2022} avoided the entrance effects by conducting experiments using a U-shaped capillary tube. Using high-speed imaging, the authors obtained detailed information on the interface shape and the dynamic contact angle.} Besides the oscillation phenomenon, also the early stage of the process has been investigated in great detail. It has been shown that, prior to the well-known Washburn regime \cite{Washburn1921} where $h \sim \sqrt{\sigma R t/\eta}$, an inertial regime exists where the rise velocity is constant \cite{Quere1997,Delannoy2019}. The work by Delannoy et al.\ \cite{Delannoy2019} used a model of the dissipation in the wedge close to the contact line to explain the enhancement of the rise velocity in the inertial regime for a pre-wetted capillary. Model extensions taking into account a dynamic contact angle have been investigated in \cite{Martic2003,Delannoy2019,Ramakrishnan2019,Wang2019}. The effect of surfactant solutions on the capillary rise is investigated in \cite{Zhmud2000}.

\paragraph{Objective and structure of this work:} We first recast the classical model by Bosanquet into a variational framework which allows to include further channels of dissipation if necessary. This is very useful in practice because further dissipation mechanisms (including dissipation due to a  dynamic contact angle \cite{Martic2003} and viscous flow close to the moving contact line \cite{Delannoy2019,Gruending2020a,Gruending2020b}) have been demonstrated to be significant for the capillary rise dynamics. The variational formulation outlined in Section~\ref{sec:model_derivation} ensures thermodynamic consistency of the models and allows a rather straightforward addition of dissipative processes based on continuum mechanical theory (briefly discussed in Section~\ref{sec:continuum-theory}). Connections between different ODE models of capillary rise become evident. The main goal of this work, addressed in Section~\ref{sec:analytical_theory}, is to generalize the analytical study of the rise height oscillations and the critical condition to a more general class of models. Using a linearization of the governing equation, we derive a generalized critical condition\footnote{More details on the mathematical properties of the model (including a proof of the well-posedness) can be found in the bachelor thesis by E.\ A.\ Ouro-Koura  (\href{https://tuprints.ulb.tu-darmstadt.de/id/eprint/24476}{https://tuprints.ulb.tu-darmstadt.de/id/eprint/24476}).}. The prediction of the theory is compared to experimental data by Qu{\'e}r{\'e} et al.\ in Section~\ref{sec:experiments}. \change{The research data for this manuscript (including the Python code to generate the results) is available in the public repository \cite{Fricke2023data} (\href{https://www.doi.org/10.5281/zenodo.8202001}{DOI:10.5281/zenodo.8202001}).}
 
\section{Continuum mechanical modeling}\label{sec:continuum-theory}
\begin{figure}[ht]
\centering
\includegraphics[width=0.5\columnwidth]{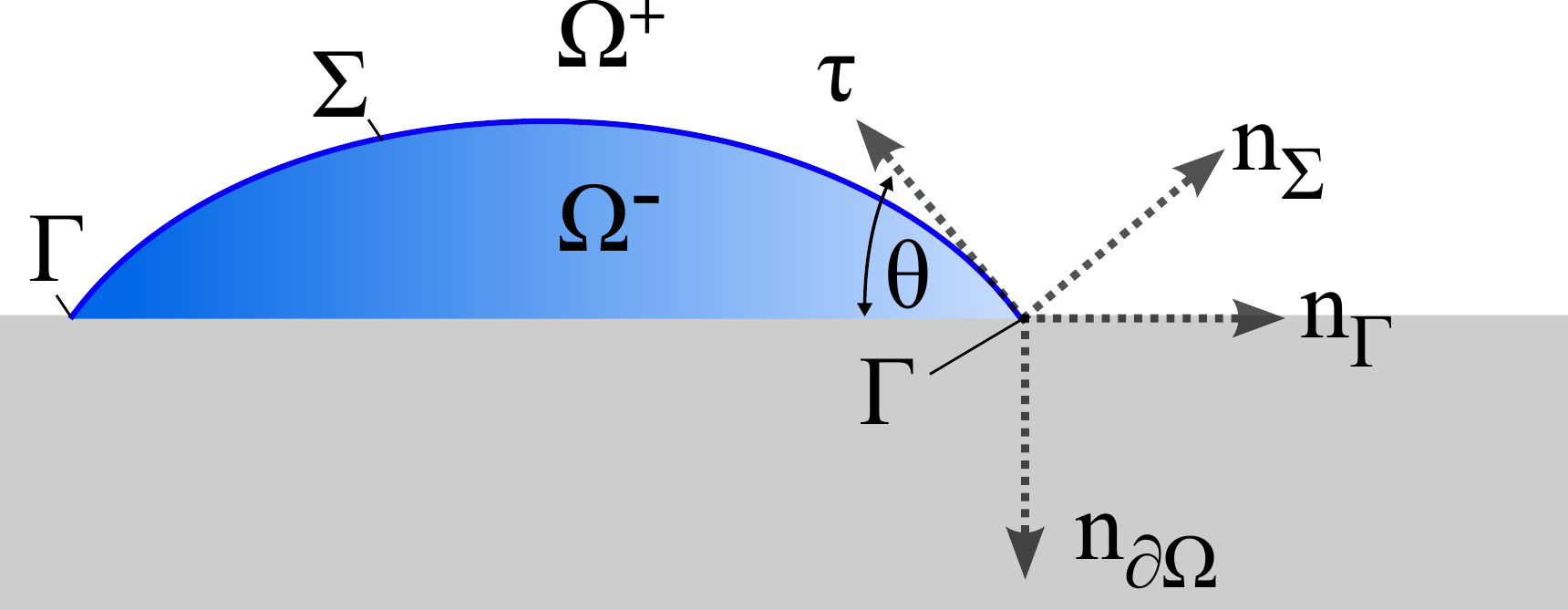}
\caption{Notation and mathematical setting.}\label{fig:notation}
\end{figure}

We briefly\footnote{For more details on the continuum mechanical modeling, we refer to Chapter~3 in \cite{Fricke2021}, \cite{Ren2007} and \cite{Bothe2016}.} recall the continuum mechanical modeling of dynamic wetting flows in the framework of the sharp interface two-phase Navier Stokes equations. The continuum mechanical models will be used to guide the derivation of ODE models, in particular, by \emph{modeling} the various dissipative processes in the system. For the sake of simplicity, we shall start with the most simple (yet non-trivial) case. We assume to have
\begin{enumerate}[(i)]
 \item \label{item:incompressible_newtonian_flow} an incompressible two-phase flow of Newtonian fluids (liquid and gas),
 \item isothermal conditions,
 \item sharp interfaces (limit of negligible interface thickness) with
 \item constant surface tensions (for liquid-gas, liquid-solid and solid-gas),
 \item no mass transfer and material interfaces,
 \item no slip for the bulk velocities at the interface and
 \item \label{item:ideal_solid_surface} impermeable and ideally homogeneous solid boundaries.
\end{enumerate}
\paragraph{Mathematical notation:} To formulate the governing equations, we shall briefly introduce the necessary notation below (see Fig.~\ref{fig:notation}). More details on the mathematical description of moving interfaces and contact lines can be found in \cite{Pruess2016} and \cite{Fricke2019}. We consider the problem in a domain $\Omega$ with (at least piecewise) smooth boundary $\partial\Omega$. The fluid interface at time $t \in I$ (for an interval $I$) is denoted as $\Sigma(t)$. In the sharp interface model, it is assumed to have zero thickness, i.e.\ it is a (smooth) hypersurface embedded in the (two or three-dimensional) domain $\Omega$. To ensure sufficient regularity in time, we also assume that
\[ \text{gr} \Sigma = \bigcup_{t \in I} \{ t \} \times \Sigma(t)  \]
is at least a $\mathcal{C}^1$-hypersurface in $\RR \times \RR^3$. The orientation of $\Sigma(t)$ is described by a continuous normal field $\nsigma(t,\cdot)$. The interface decomposes the domain into disjoint parts $\Omega^\pm(t)$, occupied by the two fluid phases (also called ``bulk phases'')
\[ \Omega = \Omega^+(t) \cup \Sigma(t) \cup \Omega^-(t). \]
We assume that each physical quantity is continuous within $\Omega^+(t)$ and $\Omega^-(t)$, respectively, with well-defined one-sided limits up to the interface. Clearly, quantities like the density will be discontinuous across the sharp interface. We define the jump of any quantity $\phi$ at a point $x \in \Sigma(t)$ as
\begin{align*}
\jump{\phi}(t,x) := \lim_{h \rightarrow 0^+} [\phi(t,x+h\nsigma) - \phi(t,x-h\nsigma).]
\end{align*}
The geometry of the boundary $\partial\Omega$ is described by the (piecewise smooth) unit outer normal field $\ndomega$. The \emph{contact line} $\Gamma(t)$ is defined as the line of intersection (if existent) of the interface with the solid boundary and denoted as
\[ \Gamma(t) := \bar\Sigma(t) \cap \partial\Omega(t). \]
At the contact line, we can define the \emph{contact angle} geometrically via the relation
\[ \cos \theta(t,x) = - \nsigma(t,x) \cdot \ndomega(t,x), \quad x \in \Gamma(t). \]
Finally, we denote by $\normalspeed \in \RR$ the speed of normal displacement of the interface. The latter is a kinematic property of the moving interface defined as
\[ \normalspeed(t,x) := \gamma'(0) \cdot \nsigma(t,x), \]
where $\gamma$ is an arbitrary $\mathcal{C}^1$ curve on the moving interface passing through the point $x \in \Sigma(t)$, i.e.\
\[ \gamma: I=(-\varepsilon, \varepsilon) \rightarrow \RR^3, \quad \gamma(s) \in \Sigma(s) \quad \forall s \in I, \quad \gamma(0) = x. \]

\paragraph{Modeling framework:} We consider a frame of reference where the boundary $\partial\Omega$ is not moving. Then, the assumptions \eqref{item:incompressible_newtonian_flow}-\eqref{item:ideal_solid_surface} lead to the following system of Partial Differential Equations describing the balance of mass and momentum in the bulk phases and at the interface as well as the interface kinematics
 \begin{equation}\label{eqn:two-phase-framework}
 \begin{aligned}
 \rho (\partial_t v + v \cdot \nabla v) - \eta \Delta v + \nabla p = \rho g, \quad \nabla \cdot v = 0, \quad &\text{in} \ \Omega\setminus\Sigma(t),\\
 \jump{v} = 0, \quad \jump{p \mathbbm{1} - S} \nsigma = \sigma \kappa \nsigma, \quad &\text{on} \ \Sigma(t),\\
 v \cdot \ndomega = 0 \quad &\text{on} \ \partial\Omega,\\
 V_\Sigma = v \cdot \nsigma \quad &\text{on} \ \Sigma(t).
 \end{aligned}
 \end{equation}
Here we used the symbols
\[ S = 2 \eta D = \eta (\nabla v + \nabla v^\transpose) \]
for the (Newtonian) viscous stress tensor (proportional to the rate-of-deformation tensor $D$), $p$ for the pressure and $\kappa := - \nabla_\Sigma \cdot \nsigma$ for the mean curvature of the interface. The system \eqref{eqn:two-phase-framework} still requires some closure relations to model both the wettability of the surface and the mobility of the contact line. 

\paragraph{Closure relations and dissipative mechanisms:} In order to derive thermodynamically consistent closure relations, we consider the free energy functional
\[ E(t) := \int_{\Omega\setminus\Sigma(t)} \frac{\rho v^2}{2} \, dV + \int_{\Sigma(t)} \sigma \, dA + \int_{W(t)} \sigma_w \, dA. \]
Here $W(t)$ denotes the wetted part of the solid boundary. It is now a purely mathematical exercise (see \cite{Ren2007} and Theorem~3.6 in \cite{Fricke2021}) to show that the rate of change of the free energy for a (sufficiently regular) solution of the system \eqref{eqn:two-phase-framework} with $g=0$ (i.e.\ no external forcing) is given as
\begin{align}
\label{eqn:energy_dissipation_semiclosed_model}
\boxed{\ddt{E} = -2 \int_{\Omega\setminus\Sigma(t)} \eta D:D \, dV + \int_{\partial\Omega} v_\parallel \cdot (S\ndomega)_\parallel \, dA + \sigma  \int_{\Gamma(t)} (\cos \theta - \cos \theta_0) \, V_\Gamma \, dl.}
\end{align}
Here $\phi_\parallel$ denotes the projection of a vector field $\phi$ to the tangent space of $\partial\Omega$, i.e.\ the ``parallel part''
\[ \phi_\parallel := \phi - (\phi \cdot \ndomega)  \, \ndomega \quad \text{at} \ \partial\Omega.  \]
Note that, assuming a partial wetting situation, we define the equilibrium contact angle $\theta_0$ in \eqref{eqn:energy_dissipation_semiclosed_model} as the unique solution of the Young equation
\[ \sigma \cos \theta_0 + \sigmawet = 0. \]
To satisfy the second-law of thermodynamics, we shall define closure relations (in this case boundary conditions) to make sure that
\[ \ddt{E} \leq 0. \]
We notice that the first integral in \eqref{eqn:energy_dissipation_semiclosed_model} (viscous dissipation in the bulk) is already dissipative for $\eta \geq 0$ (since it is a quadratic form in $D$ with the correct sign in front). So essentially, we are looking for closure relations to treat the other two integrals ($\int_{\partial\Omega} \dots$ and $\int_{\Gamma} \dots$). The classical approach is to consider the two integrals separately (see below). Interestingly, the ``Generalized Navier Slip Condition (GNBC)'' can be understood as an attempt to derive a closure relation directly for the sum of the two integrals (see Section~3.4 in \cite{Fricke2021} for more details).

\begin{enumerate}[(i)]
 \item \textbf{No-slip and Slip boundary conditions:} Obviously, the second integral \eqref{eqn:energy_dissipation_semiclosed_model} vanishes if the usual no-slip condition is applied. However, as pointed out before, this approach does not allow for a moving contact line. In fact, it has been shown by Huh and Scriven \cite{Huh1971} that the first integral becomes singular for a hypothetical moving contact line with no-slip (in the present modeling framework).\\
 \\
 A standard approach in continuum thermodynamics \cite{Groot1984} is to apply \emph{linear} relations between thermodynamic ``forces'' and ``fluxes'' to guarantee a dissipative process. This leads to quadratic forms in the entropy production. In this case, the linear closure relation reads as 
\begin{align}\label{eqn:navier_slip}
 -\beta v_\parallel = (S \ndomega)_\parallel \quad \text{at} \ \partial\Omega \quad \text{with} \quad \beta \geq 0.
\end{align}
The latter equation is nothing but the well-known Navier slip condition. Notably, the term $-\beta v_\parallel$ models a \emph{friction} force opposing the fluid motion along the boundary and balanced by the tangential viscous stress component. The parameter $\beta$ plays the role of a friction coefficient. Employing \eqref{eqn:navier_slip}, the second integral in \eqref{eqn:energy_dissipation_semiclosed_model} reads as
\[ \int_{\partial\Omega} v_\parallel \cdot (S\ndomega)_\parallel \, dA = - \int_{\partial\Omega} \beta v_\parallel^2 \, dA \leq 0. \]
We note that the quantity
\begin{align}\label{eqn:slip_length_definition}
L := \frac{\eta}{\beta}
\end{align}
has the dimension of a length and is called the "slip length". It is an important parameter in this model, controlling the mobility of the contact line at a given shear rate. Evidently, it has a strong influence on the contact line dynamics. The Huh and Scriven paradox is recovered in the limit $L \rightarrow 0$ (or, equivalently, $\beta \rightarrow \infty$).
\item \textbf{Dissipation at the contact line:} Another way for the system to dissipate energy is through a dynamic contact angle. The \emph{contact line dissipation} is described by the third integral, i.e.\
\[\sigma  \int_{\Gamma(t)} (\cos \theta - \cos \theta_0) \, V_\Gamma \, dl. \]
Clearly, the contact line dissipation is zero if the contact angle is prescribed to be equal to the equilibrium value. Another possibility is to employ a linear closure relation (similar to the one for the slip condition). A linear closure relation for the contact line dissipation (also known as the linear response theory for the dynamic contact angle) reads as
\begin{align}\label{eqn:linear_contact_angle_closure}
- \zeta \clspeed = \sigma (\cos \theta - \cos \theta_0) \quad \text{at} \ \Gamma(t) \quad \text{with} \quad \zeta \geq 0. 
\end{align}
Interestingly, also equation~\eqref{eqn:linear_contact_angle_closure} can be understood as a balance of a ``contact line friction'' force $-\zeta \clspeed$ with a stress force, in this case the (out-of-balance) Young stress $\sigma (\cos \theta - \theta_0)$. In this sense, equation~\eqref{eqn:linear_contact_angle_closure} is quite similar to \eqref{eqn:navier_slip}. Moreover, we note that \eqref{eqn:linear_contact_angle_closure} is equivalent to the small contact line speed limit of the Molecular Kinetic Theory of wetting (see Section~\ref{sec:martic_model_section}).\newline
\newline
Clearly, also non-linear closure relations for the contact angle are possible. For example, the molecular kinetic theory predicts that
\begin{equation}
\clspeed \sim \sinh \left[\frac{\sigma \left(\cos \theta_0-\cos \theta \right)}{2 n k_B T}\right].
\end{equation}
From the perspective of the entropy principle, any functional relation between the capillary number and the contact angle of the form
\[ \Ca = g(\cos \theta_0 - \cos \theta) \quad \text{or} \quad \cos \theta_0 - \cos \theta = f(\Ca) \]
is thermodynamically consistent if the function $f$ or $g$ satisfies the inequality
\[ s g(s) \geq 0, \quad s f(s) \geq 0. \]
Any non-linear closure relation of this type may be incorporated to model the contact line dissipation $\mathcal{D}_\Gamma$ in the present modeling framework. For small velocities of the contact line, they will be well approximated by a linear relation of type \eqref{eqn:linear_contact_angle_closure}.
\end{enumerate}
So, in summary, we have shown that the system \eqref{eqn:two-phase-framework} together with closure relations for slip and wettability (e.g., equations \eqref{eqn:navier_slip} and \eqref{eqn:linear_contact_angle_closure}) forms a thermodynamically consistent continuum mechanical model. This modeling framework is one of the ``standard models'' in the literature, even though there is still a scientific debate about its physical validity on small scales. In fact, it is known that solutions are always at least weakly singular at the contact line (provided that $L$ is finite at the contact line); see \cite{Huh1977,Shikhmurzaev2006} and \cite{Fricke2019}. The dissipation on the right-hand side of \eqref{eqn:energy_dissipation_semiclosed_model} may serve as a guideline to \emph{model} the dissipation of the system in complexity-reduced (i.e.\ ODE based) models. We will follow this approach to classify different existing models for the capillary rise with respect to the considered dissipative processes (see Section~\ref{sec:model_derivation}).

\section{Derivation of complexity reduced models}\label{sec:model_derivation}
In the following, we will derive (using some simplifying assumptions) the following evolution equation for the dimensionless height $H=h/h_0$ as a function of dimensionless time $s=t/\tau$ with $\tau=\sqrt{h_0/g}$:
\begin{align}\label{eqn:generic_ode}
\boxed{(H H')' - H'^2 \hs(H') + H - 1 = \frac{1}{H'} \, \frac{\mathcal{D}}{\dref}.}
\end{align}
Here, the symbol $\hs$ denotes the function
\begin{align*}
\hs(x) = \begin{cases} 1 & \text{if} \ x \leq 0,\\ 0 &\text{if} \ x > 0. \end{cases} 
\end{align*}
The quantity
\[ \dref = -2\pi R \sigmawet \frac{h_0}{\tau} = - 2 \pi R \sigmawet \sqrt{h_0 g} \]
is the reference dissipation rate. It is given as the rate of release of potential wetting energy if the column rises at a typical speed $h_0/\tau = \sqrt{h_0 g}$. It is non-negative if $\sigmawet \leq 0$, i.e.\ if $h_0 \geq 0$. The idea is then to \emph{model} the various dissipation processes, i.e.\ the term $\mathcal{D}$ based on the continuum mechanical dissipation described in equation~\eqref{eqn:energy_dissipation_semiclosed_model}.\\
\\
We derive the model by simplifying the geometry considerably. Neglecting the details of the meniscus shape, the liquid volume and mass within the cylindrical column is approximated as
\[ V(t) = \pi R^2 (h_a(t) + \hat{h}) =: \pi R^2 h(t), \quad M(t) = \rho V(t) = \pi \rho R^2 h(t), \]
where $h_a$ is the ``apex height'', i.e.\ the height measured at the lowest point of the meniscus and $\hat{h}$ is a correction term that takes into account the additional mass and potential energy of the liquid close to the meniscus. It has been shown that $\hat{h}$ can be significant (depending on the physical parameters) \cite{Gruending2020b}. The correction assumes a spherical cap shape of the interface and depends on the contact angle. Mathematically, the correction reads as\footnote{The formula is also provided in \cite{Ramakrishnan2019} with reference to publications by Verschaffelt (1919) and Dorsey \cite{Dorsey1926}.}
\begin{equation}\label{eqn:rise_height_correction}
\hat{h}=R\left(\frac{3 \cos^2 \theta-2\left(1-\sin^3 \theta\right)}{3 \cos^3 \theta}\right).
\end{equation}
We will work with $h = h_a + \hat{h}$ in the following. The corrected apex height $h_a$ can be obtained from \eqref{eqn:rise_height_correction} if needed.

\paragraph{Modeling the free energy:} We follow an approach motivated by the thermodynamic considerations in Section~\ref{sec:continuum-theory} (see also \cite{Wang2019} for a similar derivation). Based on the simplified expression for the mass, we define the kinetic energy
\begin{align}
E_k(t) = \frac{1}{2} M(t) \dot{h}(t)^2 &= \frac{\pi}{2} \rho R^2 h(t) \dot{h}(t)^2.
\end{align}
The surface energy due to wetting (here we also neglect the details of the meniscus)  is given as
\[ E_w = 2 \pi R \sigmawet h(t). \]
The surface energy of the liquid-gas interface is approximated as
\[ E_\Sigma = \pi R^2 \sigma \]
and does not depend on the rise height. The gravitational energy is given as
\[ E_g = \frac{1}{2} \pi R^2 \rho g h^2. \]
Clearly, the stationary rise height can readily be found from minimizing the sum of gravitational energy and the surface energy due to wetting. The result is the well-known formula for the stationary height \eqref{eqn:jurins_height}. We will use $h_0$ in the following as the length scale for nondimensionalization. 
Following the second law of thermodynamics, the rate of change of the total free energy
\[ E = E_k + E_w + E_\Sigma + E_g \]
is given by the total dissipation in the system and, hence, $\dot{E} \leq 0$ must hold.

\paragraph{Dissipation due to entrance effects:} Since our balance volume is not closed, we have to account for the flux of kinetic energy across the bottom of the capillary. Assuming, that the fluid is flowing in with a velocity $\dot{h}$, we approximate it as
\[ \Delta E_k^{in} = \frac{1}{2} (\Delta m) \dot{h}^2, \]
where $\Delta m = \rho \pi R^2 \dot{h}(t) \Delta t$ is the amount of mass transferred within time $\Delta t$. Consequently, the inflow rate of kinetic energy is modeled as
\begin{align}
\dot{E}_k^{in} = \frac{1}{2} \rho \pi R^2 \dot{h}^3.
\end{align}
There is a variety of models discussed in the literature regarding the dissipation at the entrance (see \cite{Lorenceau2002,Wang2019}). It is argued that eddies appear at the entrance of the tube because of the abrupt contraction between the tank and the tube. We do not go into details but follow the model used in \cite{Quere1999,Lorenceau2002}. Mathematically, this model states that for the rising column, the rate of inflow of kinetic energy causes an equally large contribution for the dissipation, i.e.
\begin{align*}
\ddt{} (E_k + E_w + E_\Sigma + E_g) = \mathcal{D} - \dot{E}_k^{in} \, \leq \, 0 \quad \text{if} \quad \dot{h} \geq 0.
\end{align*}
Conversely, if the liquid column is falling, there is an analogous term in the dissipation with a reversed sign leading to
\begin{align*}
\ddt{} (E_k + E_w + E_\Sigma + E_g) = \mathcal{D} + \dot{E}_k^{in} \, \leq \, 0 \quad \text{if} \quad \dot{h} \leq 0.
\end{align*}
To summarize the two cases, we may write (using the signum function)
\begin{align*}
\ddt{} (E_k + E_w + E_\Sigma + E_g) = \mathcal{D} - \dot{E}_k^{in} \sgn(\dot{h}) \, \leq \, 0.
\end{align*}
Note that the term $- \dot{E}_k^{in} \sgn(\dot{h}) \leq 0$ always describes a dissipative process. It is convenient to rewrite the equation using
\[ \ddt{} E_k + \dot{E}_k^{in} \sgn(\dot{h}) = \frac{\pi}{2} \rho R^2 (\dot{h}^3 + 2 h \dot{h} \ddot{h} + \dot{h}^3 \, \sgn(\dot{h})) = \pi \rho R^2 \dot{h} \left( \ddt{}(h\dot{h}) + \dot{h}^2 \frac{\sgn(\dot{h})-1}{2} \right). \] 
Noticing that $\hs(\dot{h})=-(\sgn(\dot{h})-1)/2$ and using that (in our model) $E_\Sigma$ is constant, we obtain
\begin{equation}
\begin{aligned}\label{eqn:generic_ode/derivation_part1}
\mathcal{D} &= \pi \rho R^2 \dot{h} \left( \ddt{}(h \dot{h}) - \dot{h}^2 \hs(\dot{h}) \right) + \ddt{}E_w + \ddt{} E_g \\
&= \pi \rho g R^2 \dot{h} \left(\frac{1}{g} \ddt{} (h \dot{h}) - \frac{\dot{h}^2}{g} \hs(\dot{h}) - h_0  + h \right).
\end{aligned}
\end{equation}
Here we used the equation for the stationary height $h_0 = -2\sigmawet/(\rho g R)$. Subsequently, equation \eqref{eqn:generic_ode/derivation_part1} is simplified by introducing the non-dimensional description $H(s) = h(s \tau)/h_0$. This yields
\begin{equation}
\begin{aligned}
\frac{\mathcal{D}}{H'} = \pi \rho g R^2 \frac{h_0^2}{\tau} \left(\frac{h_0}{\tau^2 g} ((H H')'-H'^2 \hs(H')) - 1 + H \right).
\end{aligned}
\end{equation}
By definition of $\tau$, we have $h_0/(\tau^2 g) = 1$ and
\[ \pi \rho g R^2 \frac{h_0^2}{\tau} = - 2 \pi R \sigmawet \frac{h_0}{\tau} =: \dref.  \]
Hence, we have derived the desired evolution equation \eqref{eqn:generic_ode}. Note that the dissipation rate $D$ is still to be modeled.

\paragraph{The inviscid case:} Assuming that no dissipation occurs ($\mathcal{D}\equiv 0$), we obtain the model
\begin{align}\label{eqn:inviscid_model}
(H H')' -H'^2 \hs(H') +  H - 1 = 0.
\end{align}
This corresponds to the model introduced in \cite{Quere1999} for the case of ``ideal rebounds''.

\subsection{Viscous dissipation in the bulk and Bosanquet's model}
We proceed by computing the viscous dissipation away from the contact line region. Assuming that the flow in the channel away from the interface satisfies the Hagen–Poiseuille equation with slip, we obtain (see Appendix~\ref{sec:appendix/bulk-dissipation} for details)
\begin{align}\label{eqn:pipeflow_dissipation}
\mathcal{D}_P = - \frac{8 \pi \eta h \dot{h}^2}{1 + \frac{4 L}{R}} = - \frac{8 \pi \eta \frac{h_0^3} {\tau^2} H H'^2}{1 + \frac{4 L}{R}}.
\end{align}
We insert the dissipation \eqref{eqn:pipeflow_dissipation} into the generic model \eqref{eqn:generic_ode} to arrive at
\begin{align}\label{eqn:model_with_pipeflow}
(H H')' - H'^2 \hs(H') + H - 1 = - \frac{\Omega}{1 + 4 L/R} \, H H',
\end{align}
where $\Omega$ is the non-dimensional parameter defined in \eqref{eqn:definition_omega}, i.e.
\[ \Omega= \sqrt{\frac{128 \eta^{2} \sigma \cos \theta_0}{R^{5} \rho^{3} g^{2}}}=\sqrt{128 \cos\theta_0} \, \frac{\mathrm{Oh}}{\mathrm{Bo}}. \]
Observe that \eqref{eqn:model_with_pipeflow} approaches the inviscid model \eqref{eqn:inviscid_model} as $L \rightarrow \infty$. On the other hand, we may neglect the contribution from the slip length $L$ in \eqref{eqn:model_with_pipeflow} if it is much smaller than the radius of the capillary. This is typically a good approximation unless the radius is very small like in nanopores. In the case of vanishing slip, equation \eqref{eqn:model_with_pipeflow} reduces to Bosanquet's model\footnote{Note that the original work by Bosanquet did not include the second term $-H'^2 \hs(H')$.} in dimensionless form, i.e.\
\begin{align}
(H H')' - H'^2 \hs(H') + \Omega \, H H' + H - 1 = 0.
\end{align}
A linearization close to stationary state shows that rise height oscillations are present in this model for $\Omega < 2$ \cite{Quere1999,Plociniczak2018}.

\subsection{Contact line dissipation and Martic's model}\label{sec:martic_model_section}
In this section, we model the contact line dissipation, i.e.\ the integral over the contact line in \eqref{eqn:energy_dissipation_semiclosed_model} given as
\[ \mathcal{D}_\Gamma =  \sigma  \int_{\Gamma(t)} (\cos \theta - \cos \theta_0) \, V_\Gamma \, dl. \]
\paragraph{Linear closure relation:} Applying a linear closure relation, we require $\clspeed$ to be proportional to the Young stress, i.e.\
\begin{equation}\label{eqn:linear_ca_model}
-\zeta \clspeed = \sigma\left(\cos \theta-\cos \theta_0\right).
\end{equation}
Here $\zeta \geq 0$ is a friction parameter with SI-unit $\text{Pa} \cdot \text{s}$ and we use the fact that $\clspeed = \dot{h}$. Note that this model agrees with the Molecular Kinetic theory (MKT) of Blake et al.\ \cite{Blake1969,Blake2002} as $\text{Ca} \rightarrow 0$. Inserting this relation yields
\[ \mathcal{D}_\Gamma = \int_\Gamma - \zeta \dot{h}^2 \, dl = - 2\pi R \zeta \dot{h}^2 = - 2\pi R \zeta \frac{h_0^2}{\tau^2} \, H'^2 \leq 0.  \]
Hence, we have the expression
\[ \frac{1}{H'} \frac{\mathcal{D}_\Gamma}{\dref} = \zeta \, \frac{h_0}{\sigmawet \tau} H' = - \frac{\zeta}{\sqrt{\sigma \rho R \cos \theta_0}} \, H' =: - \tilde{\zeta} H', \]
where
\begin{align}\label{eqn:dimensionless_cl_friction}
\tilde{\zeta} = \frac{\zeta}{\sqrt{\sigma \rho R \cos \theta_0}}
\end{align}
is the dimensionless friction coefficient. By combining the contact line dissipation $\mathcal{D}_\Gamma$ and the bulk dissipation $\mathcal{D}_P$ from the previous section (putting $L=0$), we arrive at the model introduced by Martic et al.\ \cite{Martic2003}
\begin{align}\label{eqn:martic_model}
(H H')' - H'^2 \hs(H') + \Omega \, H H' + H - 1 = - \tilde{\zeta} \, H'.
\end{align}
It is important to note that the contact line dissipation introduces an additional term proportional to $H'$. This term has a different mathematical structure than the terms appearing in \eqref{eqn:model_with_pipeflow}. Moreover, the above model is characterized by \emph{two} non-dimensional parameters, namely $\Omega$ and $\tilde{\zeta}$. This is an important qualitative difference to the classical model \eqref{eqn:bosanquet_model_dimless}.

\begin{rmk}[Viscous dissipation close to the contact line]\label{remark:viscous_dissipation_close_to_contactline}
Interestingly, the models introduced by Gründing \cite{Gruending2020a} and Delannoy et al.\ \cite{Delannoy2019} show a similar mathematical structure to the one by Martic et al.\ even though the physical mechanisms are quite different. Gründing assumes a constant contact angle but includes a model for the viscous dissipation in the vicinity of the moving contact line. The latter is computed using a known asymptotic solution of the Stokes equations for the Navier slip boundary condition. Notably, Gründing obtains an additional term $\sim - (\eta/L) \dot{h}$. Hence, Gründing's model falls into the same category as \eqref{eqn:martic_model}. This indicates that one must be careful in the mathematical modeling of the process. ``Lumping'' the effect of the viscous dissipation in the contact line vicinity into the parameter $\zeta$ (or vice versa lumping $\zeta$ into the slip length $L$) is formally possible while the modeled physical phenomenon is quite different. Delannoy et al.\ compute a wedge friction force $F_w$ using a scaling relation for the shear stress. This force is proportional to $\dot{h}/\theta$. Following the Cox-Voinov-Tanner law, it is further assumed that $\dot{h} \sim \theta^3$ leading to $F_w \sim \dot{h}^{2/3}$. 
\end{rmk}
 
\section{Mathematical theory for rise height oscillations}\label{sec:analytical_theory}
In the following, we generalize the theory of Qu{\'e}r{\'e} and P{\l}ociniczak et al.\ to the more general class of models described above. In particular, we study initial value problems of the form
\begin{equation}\label{eqn:general_ode_model}
\begin{aligned}
(H H')' + \Omega H H' - H'^2 \hs(H') + H - 1 = - \beta H',\\
\quad H(s_0) = H_0, \quad H'(s_0)=V_0.
\end{aligned}
\end{equation}
The following transformation is useful to simplify the calculations
\begin{align*}
z(s) := H(s)^2 \quad \Rightarrow z'(s) = 2 H H'(s), \quad z''(s) = 2 (H H')'.
\end{align*}
Then, equation \eqref{eqn:general_ode_model} becomes
\begin{equation}\label{eqn:z-equation}
\begin{aligned}
z''  + \left( \Omega + \frac{\beta}{\sqrt{z}} \right) z' - \frac{z'^2}{4z} \hs(z') + 2(\sqrt{z} - 1) = 0,\\
z(s_0) = H_0^2,\quad z'(s_0)=2H_0 V_0.
\end{aligned}
\end{equation}
\paragraph{Linearization close to stationary state:} In order to find the critical condition, we look at perturbations of the stationary state, i.e.\ we take
\begin{align}
\delta(s) := z(s) - 1 = H(s)^2-1.
\end{align}
Rewriting \eqref{eqn:z-equation} for $\delta$ yields the non-linear problem
\begin{align*} 
\delta''  + \left( \Omega + \frac{\beta}{\sqrt{1+\delta}} \right) \delta' - \frac{\delta'^2}{4(1+\delta)} \hs(\delta')  + 2(\sqrt{1+\delta} - 1) = 0,\\
\delta(s_0) = H_0^2 - 1, \quad \delta'(s_0)=2H_0 V_0. 
\end{align*}
We linearize this equation close to the stationary state ($\delta=0=\dot\delta$) and obtain the linear problem
\begin{align}\label{eqn:linear-epsilon-equation}
\delta''_l +  2 \xi \delta'_l + \delta_l = 0, \quad \delta_l(s_0) = H_0^2 - 1, \quad \delta'_l(s_0) = 2 H_0 V_0.
\end{align}
where $\xi = \frac{\Omega + \beta}{2} \geq 0$ is called the ``damping ratio''. 
We observe that \eqref{eqn:linear-epsilon-equation} is the equation of a damped harmonic oscillator with damping ratio $\xi$ and angular frequency $\omega_0 = 1$. As it is well-known, \change{the} equation can be solved using \change{Ansatz} functions of the form $\delta_l(s) = e^{\lambda s}$. The eigenvalues $\lambda_1$ and $\lambda_2$ are the roots of the characteristic polynomial
\[ p(\lambda) = \lambda^2 + 2 \xi \lambda + 1 = (\lambda + \xi)^2 + 1 - \xi^2. \]
Oscillations are present if the two eigenvalues are not on the real axis, i.e. if
\[ \xi < 1. \]
Hence, the critical condition for oscillations of \eqref{eqn:linear-epsilon-equation} can be expressed as
\begin{align}\label{eqn:critical_condition}
\boxed{\Omega + \beta < 2.}
\end{align}
So, as expected, the presence of the dissipative mechanism (or mechanisms) encoded in the parameter $\beta$ shifts the critical $\Omega$ to smaller values compared to the theory by Qu{\'e}r{\'e}.

\begin{rmk}[Critical condition for Martic's model]
Applied to the model by Martic et al., the critical condition \eqref{eqn:critical_condition} takes the specific form
\begin{align}\label{eqn:critical_condition_martic}
\sqrt{\frac{128 \eta^{2} \sigma \cos \theta_0}{R^{5} \rho^{3} g^{2}}} + \frac{\zeta}{\sqrt{\sigma \rho R \cos \theta_0}} < 2,  
\end{align}
where $\zeta$ is the coefficient of the contact line friction. Based on molecular dynamics simulations, an empirical relation for the friction coefficient was proposed in the form \cite{Blake2015}
\[ \zeta = \frac{n v_L \eta}{\lambda} \exp\left( \frac{\sigma(1 + \cos \theta_0)}{n k_B T} \right). \]
Here, $n$ denotes the number of adsorption sites per unit area, $\lambda$ is the average distance of a molecular jump and $v_L$ is the molecular flow volume of the liquid. As usual, $k_B$ is the Boltzmann constant and $T$ the absolute temperature. In particular, we see that $\zeta$ is proportional to the viscosity. Consequently, the left-hand side of \eqref{eqn:critical_condition_martic} is proportional to the viscosity. Moreover, we see that there is a highly non-trivial dependence on the equilibrium contact angle $\theta_0$.
\end{rmk}

\subsection{Oscillatory case}
For the case of under-critical damping, i.e.\
 \[ \xi = (\Omega + \beta)/2 < 1, \]
 we can write the general solution of the linearized equation \eqref{eqn:linear-epsilon-equation} in the form
 \begin{align}\label{eqn:general_solution_oscillatory}
 \boxed{\delta_l(s) = A \, e^{- \xi s} \cos\left(\omega (s-s_0) + \phi \right), \quad A, \ \phi, \ s_0 \in \RR,}
 \end{align}
 where
 \begin{align}
 \omega = \sqrt{1-\xi^2} = \sqrt{1-(\Omega+\beta)^2/4}.
 \end{align}
 is the oscillation frequency and
 \[ S = \frac{2 \pi}{\omega} = \frac{2\pi}{\sqrt{1-(\Omega+\beta)^2/4}} \]
 is the corresponding (non-dimensional) oscillation timescale.  We recognize that the oscillation timescale goes to infinity as $\xi$ approaches unity. Therefore, the oscillations are hardly visible close to critical damping. One can also solve the initial value problem
 \[ \delta_l(s_0) = \delta_0, \quad \delta'_l(s_0) = \dot\delta_0  \]
 to determine the phase constant $\phi$ and the ``amplitude''\footnote{Note that the parameter $A$ should not be confused with the maximum value of the function~$\delta_l(s)$.} $A$ in \eqref{eqn:general_solution_oscillatory}. Using the general solution, one obtains the system of equations
 \begin{equation*}
 \begin{aligned}
 A e^{-\xi s_0} \cos\phi = \delta_0\\
 - \xi \delta_0 - A e^{-\xi s_0} \sqrt{1-\xi^2} \sin \phi = \dot\delta_0
 \end{aligned}
 \end{equation*}
 with the solution
 \begin{align}
 A = \delta_0 e^{\xi s_0} \sqrt{1 + \frac{(\xi + \dot\delta_0/\delta_0)^2}{1-\xi^2}}
 \end{align}
 for the amplitude and
 \begin{align}
 \phi = \arctan\left(  - \frac{\xi + \dot\delta_0/\delta_0}{\sqrt{1-\xi^2}} \right)
 \end{align}
 for the phase constant. This characterizes the solution of the linear problem in the oscillatory case completely.

\subsection{Overdamped case}\label{sec:theory/highDamping}
In the case of large damping, i.e.\ 
 \[ \xi >  1, \]
 we can write the general solution of \eqref{eqn:linear-epsilon-equation} as
 \begin{align}\label{eqn:overdamped_solution}
 \delta_l(s) = A \exp([-\xi - \sqrt{\xi^2 - 1}] s) + B \exp([-\xi + \sqrt{\xi^2 - 1}]s).
 \end{align}
 In this case, both eigenvalues 
 \[ \lambda_1 = -\xi - \sqrt{\xi^2 - 1} < \lambda_2 = -\xi + \sqrt{\xi^2 - 1} < 0 \]
 are real and negative. We note that the solution \eqref{eqn:overdamped_solution} can cross the equilibrium position $\delta=0$ at most once. 
 \paragraph{Asymptotics for very large damping:} It is interesting to study the asymptotics for large $\xi \gg 1$. In fact, it is easy to show that
 \[ \lambda_1 \sim - 2 \xi, \quad \lambda_2 \sim -\frac{1}{2\xi} \quad \text{as} \quad \xi \rightarrow \infty.  \]
 Hence, in this case, the solution has the form
 \begin{align} \label{eqn:asymptotics_large_damping}
 \delta_l(s) = A \exp(- 2 \xi s) + B \exp\left(- \frac{s}{2\xi} \right). 
 \end{align}
 If $\xi$ is large, then the first part of the above solution will decay quickly and
 \begin{align}\label{eqn:asymptotics_large_damping_v2} 
 \delta_l(s) \approx \delta_0 \exp\left(- \frac{s}{2\xi} \right) 
 \end{align}
 is a good approximation to the solution. In particular, we see that the characteristic timescale of the approach to stationary state is given as $S = 2 \xi$.
 
 \subsection{Case of critical damping}
The two eigenvalues $\lambda_1$ and $\lambda_2$ degenerate to a single eigenvalue $\lambda = - \xi = -1$ in the case of critical damping, i.e.\
  \[ \xi = 1.   \]
 In this case, the general solution reads as
\[ \delta_l(s) = e^{-s} (A + B s), \quad A, B \in \RR. \]

\section{Comparison with experimental data}\label{sec:experiments}
\begin{table}[hb]
\centering
\begin{tabular}{|c|c|c|c|}
\hline
 & Ether & Ethanol & Silicon Oil \\
\hline
$\rho$ $[\text{kg}/\text{m}^3]$ & 710 & 780 & 980  \\
\hline
$g$ $[\text{m}/\text{s}^2]$ & 9.81 & 9.81 & 9.81  \\
\hline
$\eta$ $[\text{Pa}\cdot\text{s}]$ & $3 \cdot 10^{-4}$ & $1.17 \cdot 10^{-3}$ & 0.5  \\
\hline
$\sigma$ $[\text{N}/\text{m}]$ & $16.6 \cdot 10^{-3}$ & $21.6 \cdot 10^{-3}$ & $21.1 \cdot 10^{-3}$  \\
\hline
$\theta_0$ $[\text{deg}]$ & 0 & 0 & 0  \\
\hline
$R$ $[\text{m}]$ & $6.89 \cdot 10^{-4}$ & $6.89 \cdot 10^{-4}$ & $4.21 \cdot 10^{-4}$ \\
\hline
$\Omega$ $[-]$ & 0.189 & 1.01 & 750  \\
\hline
\end{tabular}
\caption{Experimental parameters \cite{Quere1997}.}\label{tab:experimental_data_quere1997}
\end{table}

\begin{figure}[ht]
\centering 
\includegraphics[width=0.7\columnwidth]{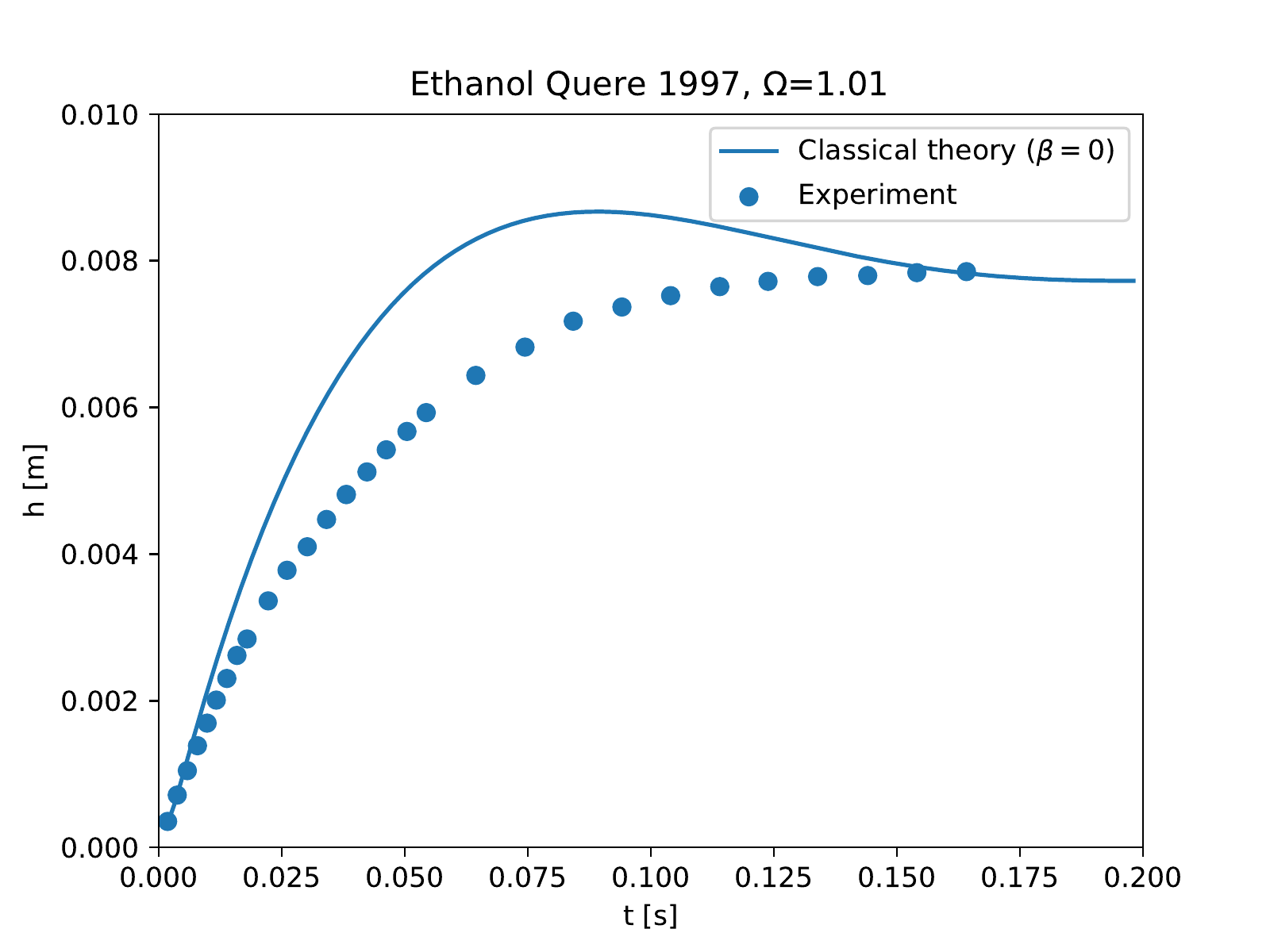}
\caption{Comparison between experimental for ethanol and the classical ODE model ($\beta=0$).}\label{fig:ether-exp-vs-classical}
\end{figure}

In the following, we apply the theory to experimental data from literature. In the well-known article from 1997 \cite{Quere1997}, Qu{\'e}r{\'e} measured the rise dynamics of three different liquids (silicon oil, ethanol and ether) in narrow glass tubes with radii below $1 \, \text{mm}$; see Table~\ref{tab:experimental_data_quere1997} for details on the experimental parameters. It was found that visible oscillations around the equilibrium height occur for low enough viscosity. Notably, the considered liquids span a wide range in terms of $\Omega$. While the viscosity of ether is very low and oscillations are visible with the naked eye, the silicon oil has a viscosity which is three orders of magnitude larger. This also causes $\Omega$ to be much larger than the critical value. Hence, a monotonic rise can be expected in this case. Ethanol lies well in between these two cases, with a $\Omega$-value close to unity. In fact, a strong oscillation is predicted in this case by the classical model (i.e.\ equation \eqref{eqn:general_ode_model} with $\beta=0$); see Figure~\ref{fig:ether-exp-vs-classical}. Even though this strong oscillation is not found in the experimental data, we will show below that this experiment lies still within the oscillatory regime. In the following, we show the experimental data in non-dimensional form, i.e.\
\[ H = h/h_0 \quad \text{and} \quad s = t/\sqrt{h_0/g} \]
where $h_0$ is the rise height in stationary state and $g$ is the gravitational acceleration.

\subsection{Experimental comparison for Ethanol}
The data for ethanol are quite interesting since $\Omega \approx 1.01$ is well below 2, but no oscillation is directly visible. From the comparison with the classical theory reported in Figure~\ref{fig:ether-exp-vs-classical}, it is obvious that additional channels of dissipation should be considered to achieve a reasonable match with the experimental data. Martic et al.\ \cite{Martic2003} studied the same dataset using the dynamic contact angle model 
\[ \zeta \clspeed = \sigma (\cos \theta_0 - \cos \theta). \]
They achieved a satisfactory fit of the data for the contact line friction
\[ \zeta = 80 \, \text{m}\text{Pa} \cdot \text{s}. \]
As pointed in equation \eqref{eqn:dimensionless_cl_friction} within Section~\ref{sec:martic_model_section}, one can convert the contact line friction into the dimensionless parameter $\beta$ according to
\[ \beta = \frac{80 \, \text{m}\text{Pa} \cdot \text{s}}{\sqrt{\sigma \rho R \cos \theta_0}} \approx \frac{80 \, \text{m}\text{Pa} \cdot \text{s}}{108 \, \text{m}\text{Pa} \cdot \text{s}} \approx 0.74. \]
We notice that for this value of $\beta$, the sum of $\Omega$ and $\beta$, i.e.\
\[ \Omega + \beta \approx 1.01+0.74 = 1.75 < 2 \]
is still below 2 and rise height oscillations are expected. However, we note that the system is rather close to the critical damping. The expected non-dimensional timescale for the oscillation from the linear theory is
\[ S = \frac{2\pi}{1-\xi^2} \approx \frac{2\pi}{1-1.75^2/4} \approx 13. \]
Hence, we cannot expect to see more than half an oscillation period within the given experimental dataset for ethanol. In order to investigate whether or not the experimental data contain oscillations, we introduce the function $\Psi$ defined as
\begin{align}\label{eqn:underdamped_psi_function}
\psi(s) := e^{\xi s} \delta(s) = e^{\xi s} (H(s)^2 - 1). 
\end{align}
By plotting the function $\psi$ based on the expected exponential decay part, we can study only the oscillatory part of the solution. However, we must be careful with the interpretation of $\psi$ for large values of the non-dimensional time $s$. This is because any error in $H(s)^2-1$ will be amplified by the exponential factor $e^{\xi s}$.\newline
\begin{figure}[ht]
\subfigure[Theory for $\beta=0.8$ vs.\ experiment.]{\includegraphics[width=0.5\columnwidth]{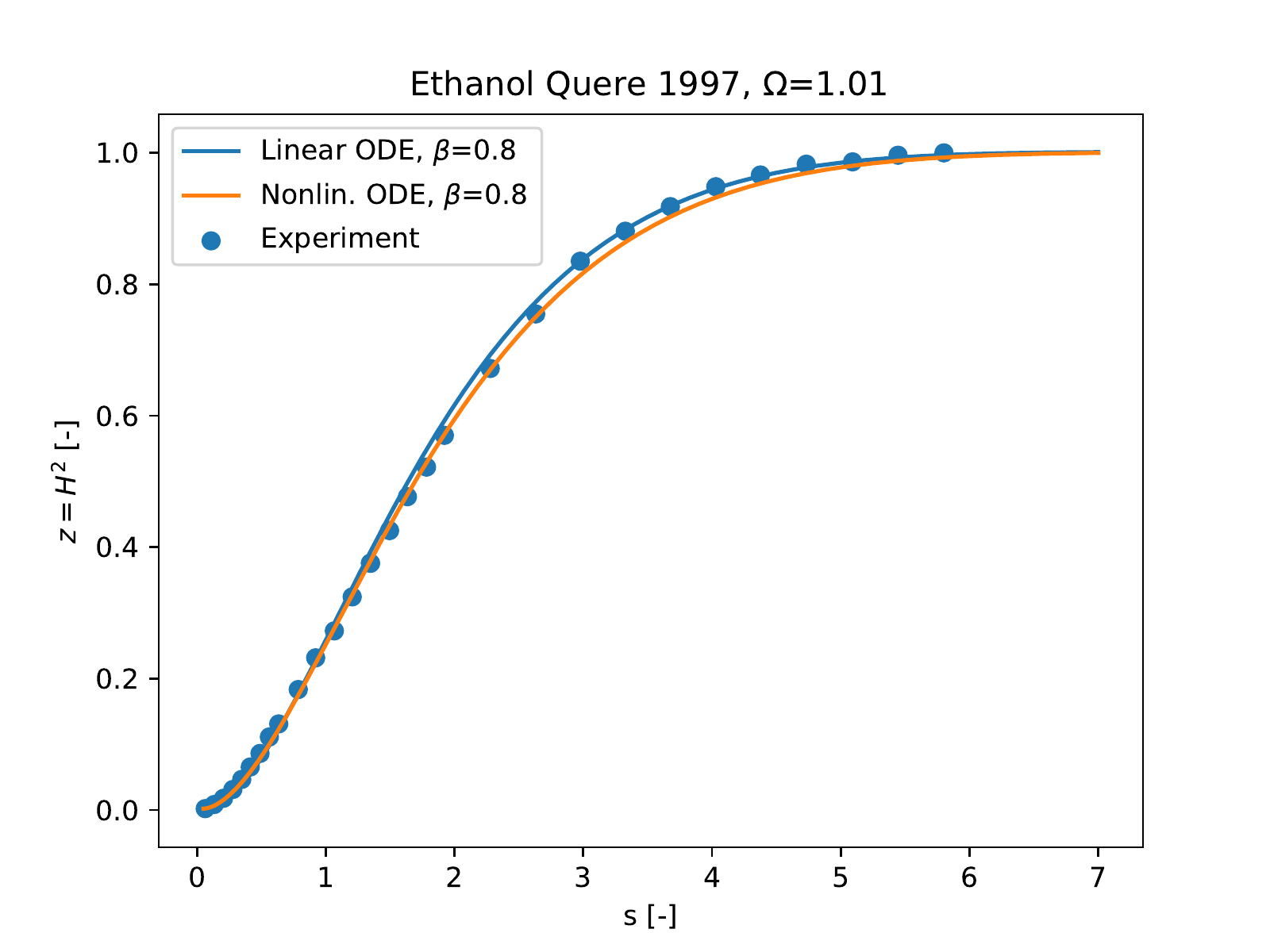}\label{fig:resuls/quere/ethanol-1}}
\subfigure[Visualization of the oscillatory part (see eq.~\eqref{eqn:underdamped_psi_function}).]{\includegraphics[width=0.5\columnwidth]{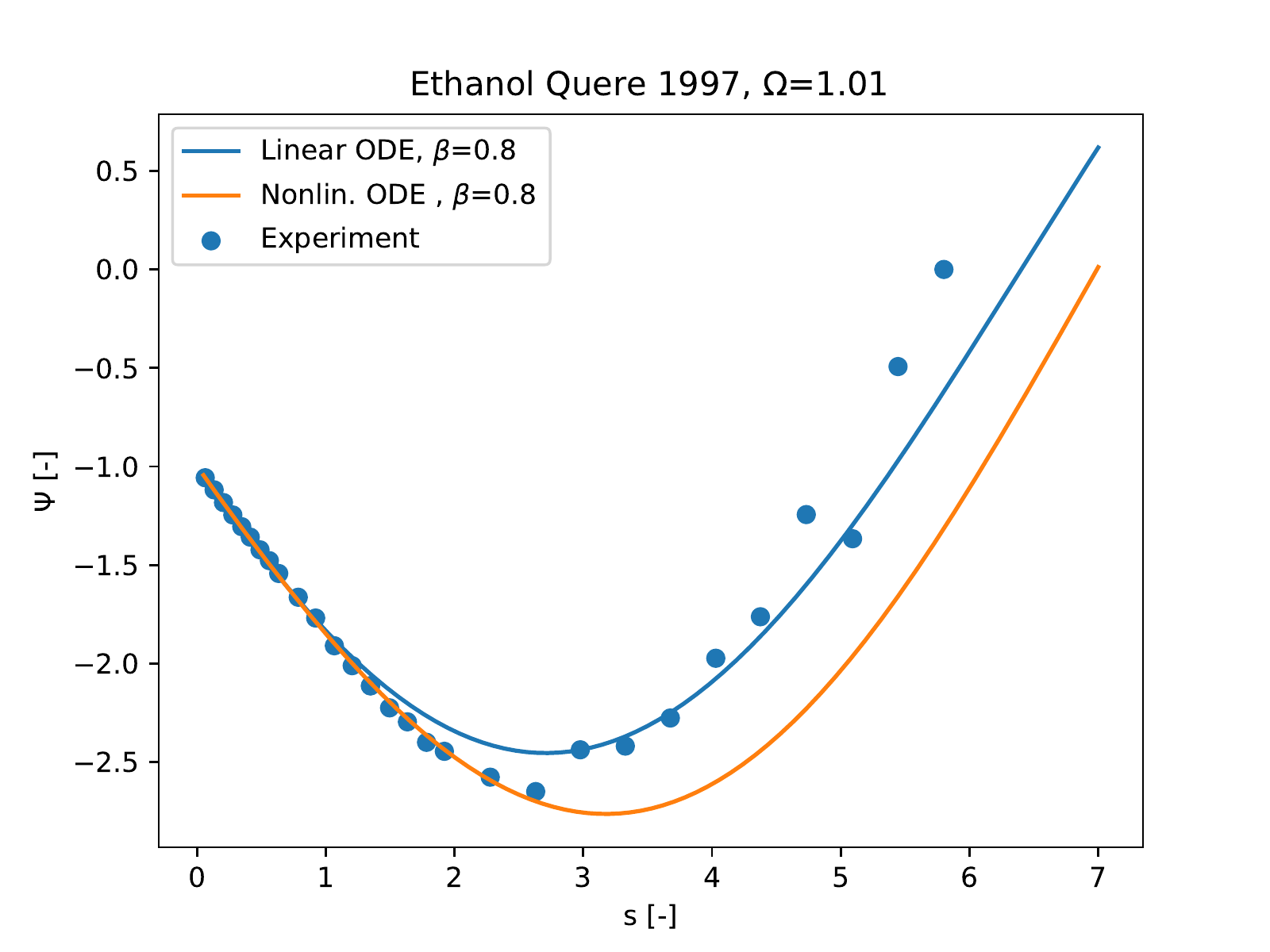}\label{fig:resuls/quere/ethanol-2}}
\caption{Experimental comparison for Ethanol in \cite{Quere1997}.}\label{fig:results_quere_ethanol}
\end{figure}

The results for Ethanol are reported in Figure~\ref{fig:results_quere_ethanol}. We compare the experimental data for $H^2$ as function of non-dimensional time $s$ with the solution of the ODE model \change{\eqref{eqn:general_ode_model}}. A good match is found for $\beta=0.8$ (slightly larger value than the one reported in \cite{Martic2003}); see Fig.~\ref{fig:resuls/quere/ethanol-1}. Notably, the existence of the oscillation in the experimental data set is confirmed by the plot of the function $\psi$ in Fig.~\ref{fig:resuls/quere/ethanol-2}. Even good quantitative agreement \change{for the frequency and amplitude} is found between the \change{ODE solution} and the experimental data for $s \lesssim 3$. \change{Notably, the solution of the linearized ODE seems to agree better with the experimental data than the solution for the non-linear equation for $s > 3$. However, this difference can hardly be deemed significant because the two solutions are quite close to each other given the typical experimental uncertainty (which is not quantified in \cite{Quere1997}). Notice also that $\psi(3) = e^{1.75 \cdot 3}(H^2-1)\approx 190 (H^2 -1)$. Hence, a relative measurement error below $0.5\%$ for $H^2$ would be necessary to evaluate the data beyond $s=3$.} \newline
\newline
The confirmation of the existence of the oscillation is remarkable since this particular experiment has been believed to be in the monotonic regime \change{because no oscillation could directly be observed in the data. This observation appeared to be in contraction with the classical theory which predicts an oscillation (since $\Omega \approx 1.01 < 2$)} \cite{Martic2003}. P{\l}ociniczak and {\'{S}}wita{\l}a noted that ``a possible reason for this discrepancy probably lies within the very high difficulty of judging which of the two behaviors take place'' when the system is close to critical damping \cite{Plociniczak2018}. Fortunately, we were able to reveal the ``hidden oscillation'' based on the analytical solution of the linearized problem.

\subsection{Experimental comparison for Ether}
\begin{figure}[hb]
\subfigure[Theory for $\beta=0.15$ vs.\ experiment.]{\includegraphics[width=0.5\columnwidth]{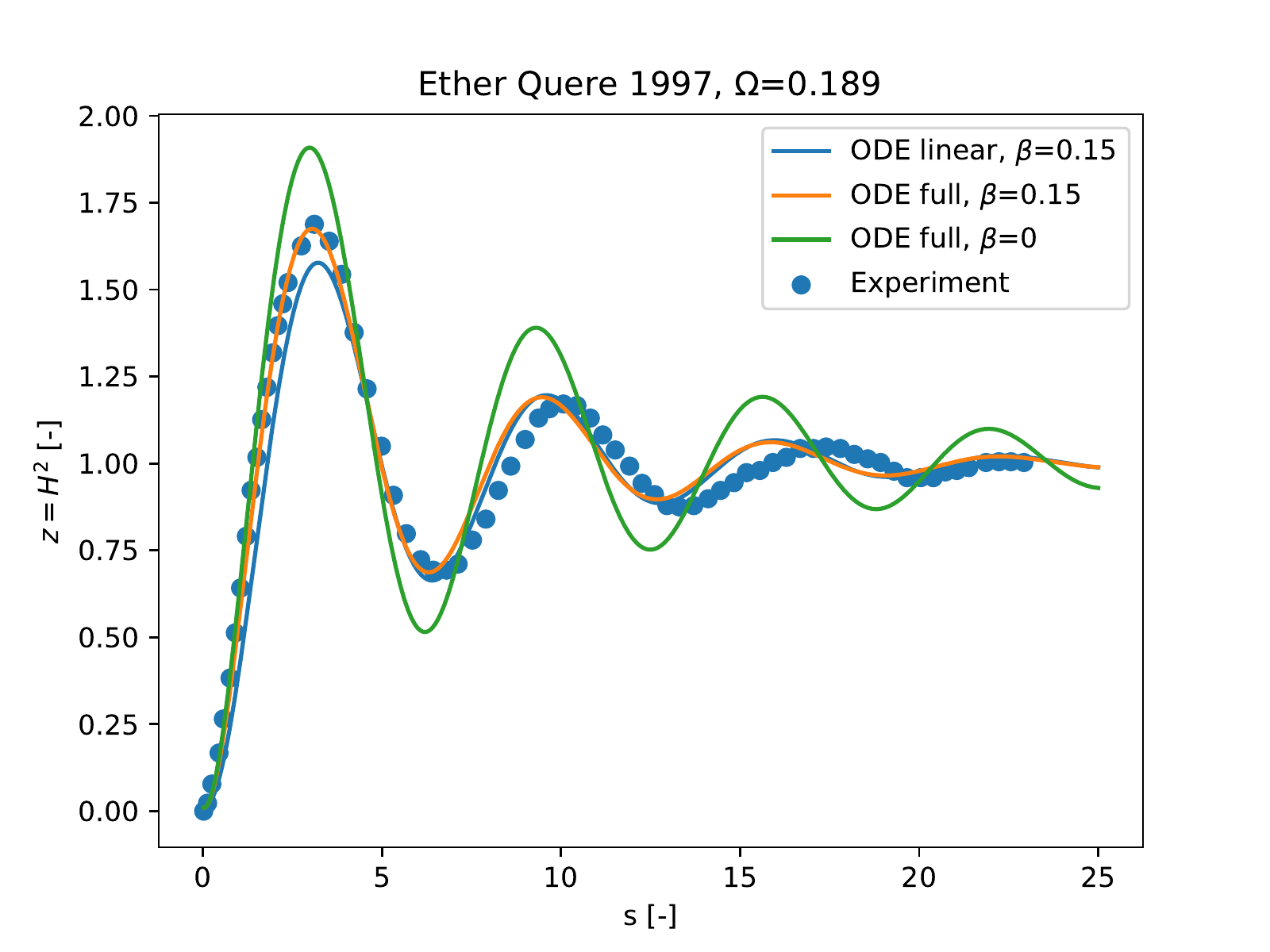}\label{fig:resuls/quere/ether-1}}
\subfigure[Visualization of the oscillatory part (see eq.~\ref{eqn:underdamped_psi_function}).]{\includegraphics[width=0.5\columnwidth]{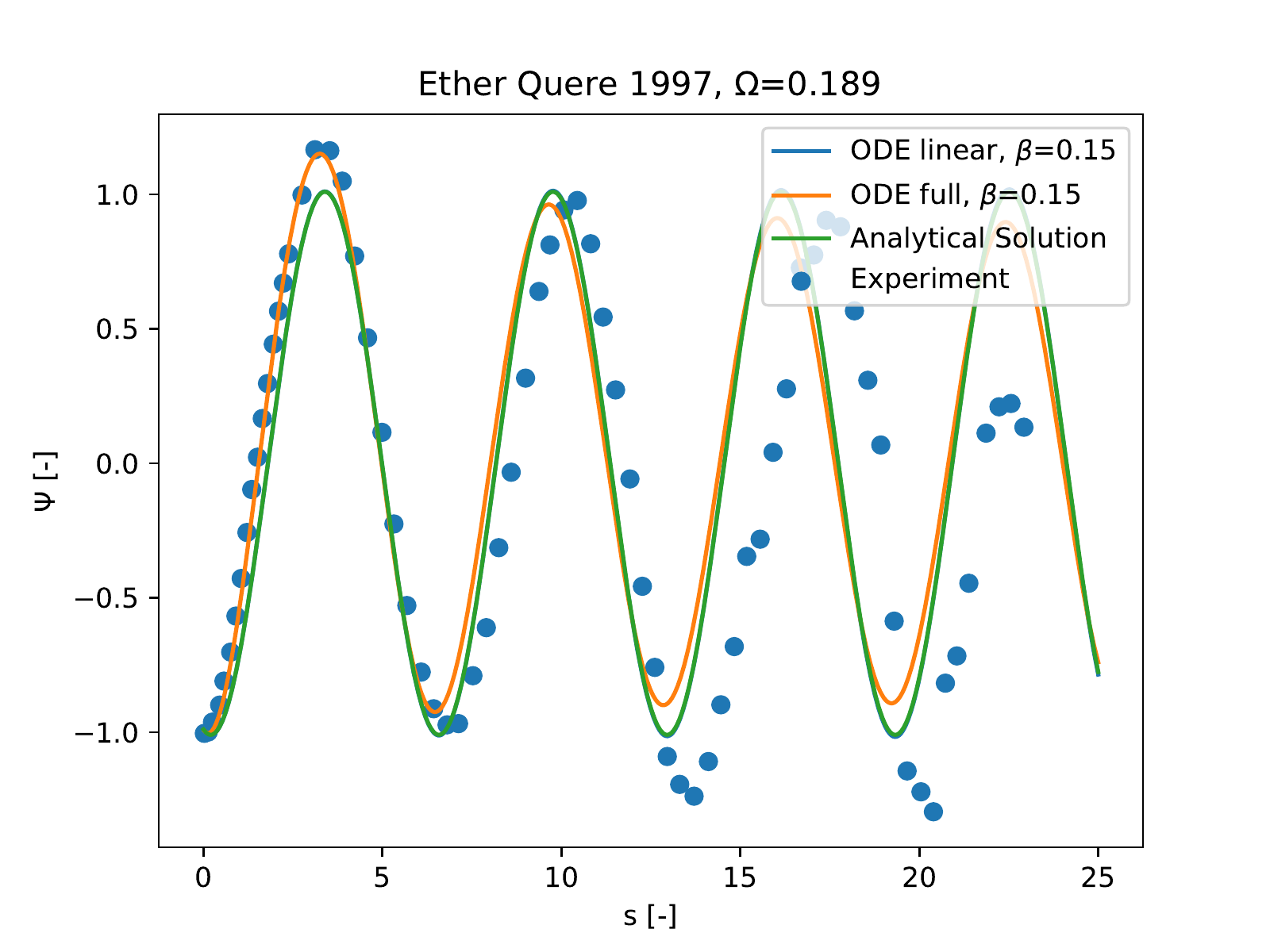}\label{fig:resuls/quere/ether-2}}
\caption{Experimental comparison for Ether in \cite{Quere1997}.}
\end{figure}
The experiment for the low viscosity liquid ether described in \cite{Quere1997} has received much attention in the literature because a strong oscillation is visible in the data (see Fig.~\ref{fig:resuls/quere/ether-1}). From the physical values, we find $\Omega \approx 0.189$. Also in this case, some additional dissipation appears to be active. Qu{\'e}r{\'e} et al.\ noted that ``the curve $\Omega=0.3$ is found to be in excellent agreement with the data (for which we rather have $\Omega=0.2$)'' \cite{Quere1999}. Indeed, we also find that the classical model ($\Omega=0.189$, $\beta=0$) overpredicts the strength of the oscillation. A good quantitative agreement (at least for the first oscillation) is found for $\beta=0.15$ which gives $\Omega+\beta \approx 0.34$ comparable to the value of $0.3$ used in \cite{Quere1999}. Plotting the function $\psi$ in Fig.~\ref{fig:resuls/quere/ether-2} allows to isolate the oscillatory part of the data and to study the oscillation characteristics in more detail. We note that already the linear equation is able to capture the essential dynamics with remarkable accuracy, given the fact that it is obtained from a linearization close to the stationary state. A deviation between the experimental data and both the linear and the non-linear model are found starting from the second  period of oscillation. Physically, this most likely means that using a simple linear dissipation model like \eqref{eqn:linear_ca_model} might be insufficient over the full trajectory of the experiment. Finally, evaporation might also play a significant role (see \cite{Ramon2008}) even though the whole process takes only $0.6 \, \text{s}$.

\subsection{Experimental comparison for Silicon Oil}
\begin{figure}[ht]
\subfigure[Asymptotic solution \eqref{eqn:asymptotics_large_damping_v2} vs. experiment ($\beta=0$)]{\includegraphics[width=0.5\columnwidth]{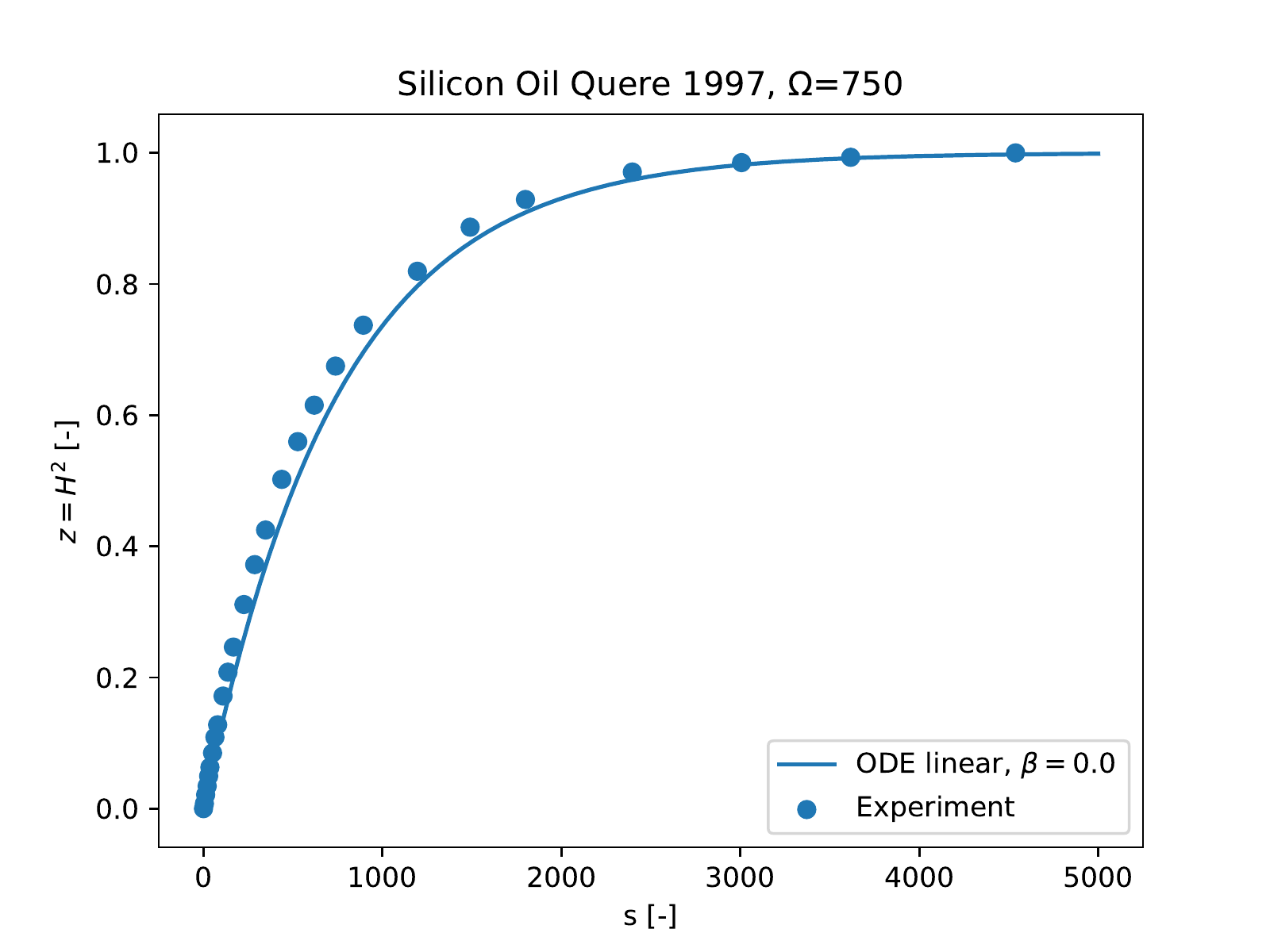}\label{fig:resuls/quere/silicon-oil-2}}
\subfigure[Plot of the function \change{$\Lambda(s)$ (see eq.\ \eqref{eqn:overdamped-lambda-function}).}]{\includegraphics[width=0.5\columnwidth]{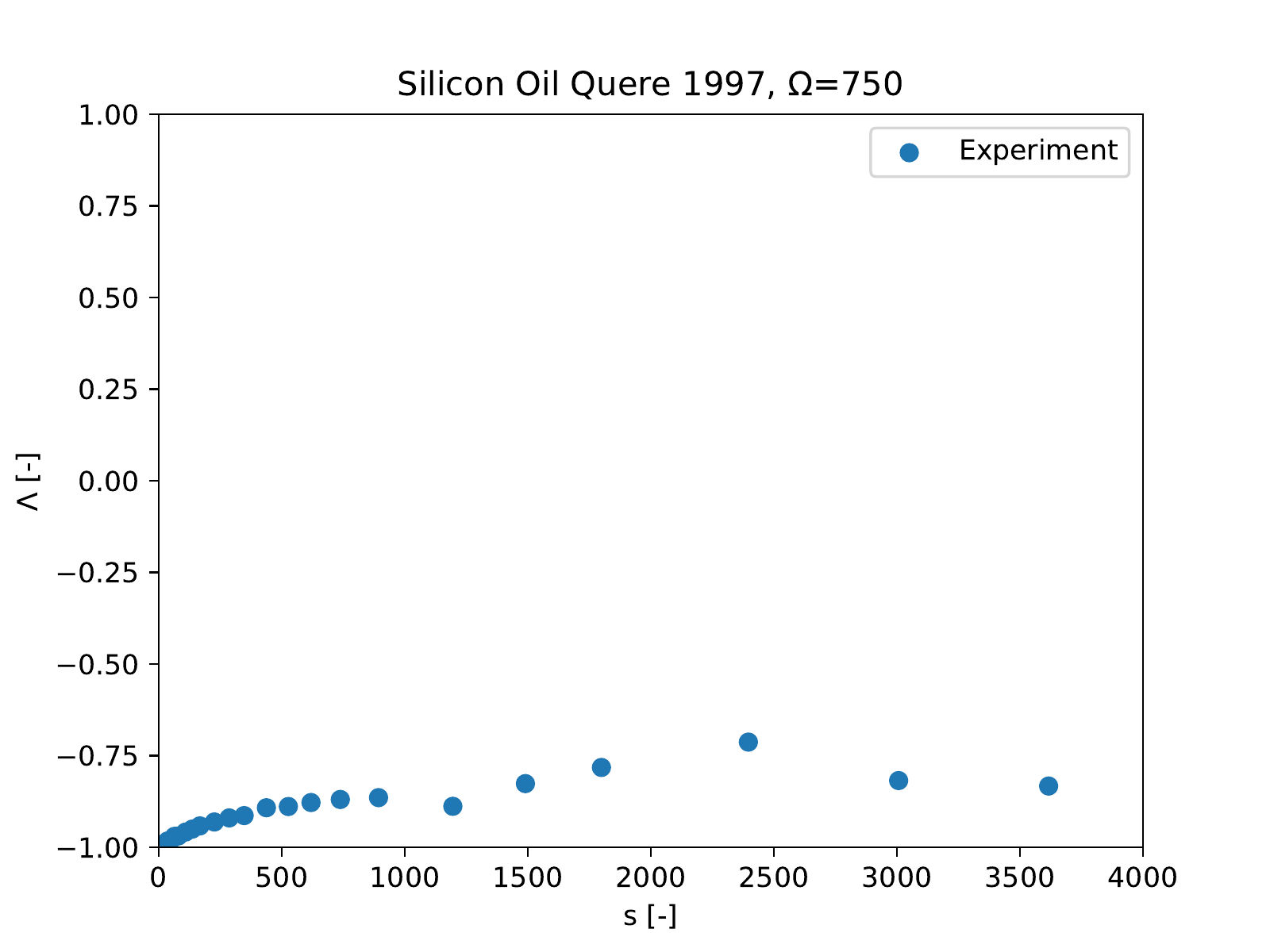}\label{fig:resuls/quere/silicon-oil-3}}
\caption{Experimental comparison for Silicon Oil in \cite{Quere1997}.}
\end{figure}
Finally, we study the case of the highly viscous silicon oil reported in \cite{Quere1997}. Since the parameter $\Omega = 750$ is very large in this case, we may describe the dynamics using the asymptotic formulas derived in Section~\ref{sec:theory/highDamping} for the case of a very high damping. Figure~\ref{fig:resuls/quere/silicon-oil-2} shows that the asymptotic solution of the linearized equation, i.e.\
\[ H(s) = \sqrt{1 + \delta_l(s)} \quad \text{with} \quad \delta_l(s) = - \exp\left(\frac{s}{750} \right) \]
shows a quite good agreement with the experimental data. Notably, $\beta=0$ is sufficient in this case. Clearly, there is no oscillatory part of the solution expected in this case. Similarly to the previous \change{definition of the function $\Psi$}, we define a function $\Lambda$ by factoring out the exponential decay part
\begin{align}\label{eqn:overdamped-lambda-function}
\Lambda(s):= \exp\left( \frac{s}{2\xi} \right) (H(s)^2 - 1). 
\end{align}
As expected, the values for $\Lambda$ are constant (aside from some minor variations caused by data uncertainties); see Fig.~\ref{fig:resuls/quere/silicon-oil-3}.
 
\section{Conclusion}
In conclusion, we have addressed the following major scientific questions in this work:
\begin{enumerate}[(i)]
\item \textbf{Mathematical Modeling:} A variational framework was outlined for the construction of complexity-reduced models. Building on the dissipation in the continuum mechanical model \eqref{eqn:energy_dissipation_semiclosed_model}, this approach offers a way to naturally extend the model by further dissipative mechanisms. We study a class of models described by equation~\eqref{eqn:general_ode_model}. The extra term of the form $\beta H'$ may have different physical origins. For example, one possible mechanism is the dissipation at the contact line caused by a (friction based) dynamic contact angle model \eqref{eqn:linear_ca_model}. Notably, also the dissipation caused by the viscous flow close to the moving contact line may lead to a term of the same mathematical form \cite{Gruending2020b} (see Remark~\ref{remark:viscous_dissipation_close_to_contactline}). This is remarkable because it indicates that there is a kind of formal equivalence between different models of the dynamic wetting process. Hence, this means that the effect of one physical process may be effectively ``lumped'' (or fitted) into a parameter of another physical process (here slip length vs.\ contact line friction). Ideally, aiming at a true \emph{prediction} of the dynamics, one should try to include all the relevant dissipative mechanisms in the model. Delannoy et al.\ \cite{Delannoy2019} used a scaling argument for the shear stress and a dynamic contact angle model to arrive at a term proportional to $(H')^{2/3}$. \change{Hence, the resulting model is not of the type \eqref{eqn:general_ode_model}. It is more involved because of the non-linear coupling between the dynamics of the contact angle and the hydrodynamic dissipation and should be studied in more detail in the future.} Asymptotic solutions of the flow close to the contact line \cite{Gruending2020a} and Direct Numerical Simulations (see, e.g., \cite{Gruending2020b,Sprittles2011b}) shall be used in the future to model additional dissipation channels. In particular, we are interested to study near equilibrium oscillations of the liquid column in detail using CFD simulations.
\item \textbf{Generalization of the mathematical theory for rise height oscillations:} The mathematical analysis by  Qu{\'e}r{\'e} is generalized to the class of models described by equation~\eqref{eqn:general_ode_model}. Following the approach by P{\l}ociniczak~et.\ al~\cite{Plociniczak2018}, we apply a substitution which allows to get rid of the non-linearity in the second-order term. A subsequent linearization close to the stationary state yields the generalization of the critical condition, i.e.\
\[ \Omega + \beta < 2, \]
where $\beta \geq 0$ is a second dimensionless parameter arising from additional dissipative processes. Moreover, we compute the full solution to the linear problem \eqref{eqn:general_solution_oscillatory} explicitly. The linear solution is remarkably effective in describing the overall dynamics (at least qualitatively) even far away from the stationary state. Despite the underlying approximation caused by the linearization, this solution may be useful for future research because we have access to all the information. Moreover, an asymptotic solution for the case of very large damping is derived \eqref{eqn:asymptotics_large_damping_v2}.
\item \textbf{Comparison with experimental data:} Finally, we revisited the experimental data by Qu{\'e}r{\'e}, i.e.\ capillary rise of Ethanol, Ether and Silicon Oil, using the generalized mathematical theory. In particular, we clarified the status of the experimental data reported for Ethanol. A strong oscillation is predicted by the classical theory since $\Omega \approx 1$. However, no oscillation is directly visible in the experimental data. A fact that has been noticed before in the literature \cite{Plociniczak2018}. With the new theory, the system is still expected to be in the oscillatory regime even though it is close to the critical damping ($\Omega + \beta \approx 1.8 < 2$). We confirmed the existence of the oscillation using a transformation of the experimental data based on the knowledge about the analytical form of the solution to the linearized problem. Indeed, an oscillation over half a period can be observed in reasonable quantitative agreement to the linearized solution. It would be interesting to study a nearly critically damped system using some more recent high resolution optical techniques in the future. Notably, also the highly damped case of silicon oil can be well-described by an asymptotic solution to the linearized problem. 
\end{enumerate}
 
\section*{Acknowledgements}
We acknowledge the financial support by the German Research Foundation (DFG) within the Collaborative Research Centre 1194 (Project-ID 265191195).

\appendix
\numberwithin{equation}{section}

\section{Viscous dissipation in the Hagen–Poiseuille flow}\label{sec:appendix/bulk-dissipation}
We now compute the viscous dissipation in the bulk far away from the contact line. Therefore, it is assumed that the flow in the bulk region below the free surface follows the Hagen–Poiseuille equation with slip boundary condition. For a given pressure gradient $G = \partial p/\partial z$, the solution in cylindrical coordinates $(r,\varphi,z)$ is given as
\begin{align}\label{eqn:3d_pipeflow}
u_z(r) = \frac{G R^2}{4 \eta} \left( 1 - \left( \frac{r}{R} \right)^2 + 2 \frac{L}{R} \right), \quad u_r = 0, \quad u_\varphi = 0.
\end{align}
In order to determine the profile for a given rise velocity, we compute the average from \eqref{eqn:3d_pipeflow}. This allows to link $G$ to $\dot{h}$. We obtain
\begin{align}\label{eqn:G_vs_dot_h}
\dot{h} = \langle u_z \rangle = \frac{1}{\pi R^2} \int_0^{2\pi} \int_0^R u_z(r) r \, dr d\varphi = \frac{G R^2}{8 \eta} \left( 1 + \frac{4L}{R} \right).
\end{align}
Moreover, the slip velocity at the wall is given as
\[ u_z(R) = \frac{GRL}{2\eta}. \]

\paragraph{Dissipation rate:} In order to compute the dissipation rate, we evaluate the first two integrals in \eqref{eqn:energy_dissipation_semiclosed_model}, i.e.
\[ -2 \int_\Omega \eta D:D \, dV \quad \text{and} \quad \int_{\partial\Omega} \inproduct{v_\parallel}{(S \ndomega)_\parallel} dA = - \int_{\partial\Omega} \lambda v_\parallel^2 \, dA\]
for the given flow profile. Computing the rate-of-deformation tensor for \eqref{eqn:3d_pipeflow} yields
\begin{align*}
D:D = \left( \frac{1}{2} \frac{\partial u_z}{\partial r} \right)^2 = \left( \frac{G r}{4\eta} \right)^2.
\end{align*}
Evaluating the integrals gives
\begin{align*}
-2 \int_\Omega \eta D:D \, dV = - 2 \eta \int_0^h \int_0^{2\pi} \int_0^R \left( \frac{G r}{4\eta} \right)^2 r \, dr d\varphi dz = - \frac{\pi h G^2 R^4}{8 \eta}.
\end{align*}
and 
\begin{align*}
- \int_{\partial\Omega} \lambda v_\parallel^2 \, dA = - \int_0^{2\pi} \int_0^h \frac{\eta}{L} \left( \frac{G R L}{2\eta} \right)^2 dz R d\varphi = - \frac{\pi h G^2 R^3 L}{2 \eta}.
\end{align*}
So the two contributions together give
\[ \mathcal{D}_P = -2 \int_\Omega \eta D:D \, dV - \int_{\partial\Omega} \lambda v_\parallel^2 \, dA = - \frac{\pi h G^2 R^4}{8 \eta} \left( 1 + \frac{4L}{R} \right). \]
We employ relation \eqref{eqn:G_vs_dot_h} to replace the pressure gradient $G$ in the above equation. This finally yields the relation
\begin{align}\label{eqn:appendix/pipeflow-dissipation}
\mathcal{D}_P = - \frac{8 \pi \eta h \dot{h}^2}{1 + \frac{4 L}{R}}
\end{align}
Note that $\mathcal{D}_P \leq 0$ provided that $h\geq0$. Notably, the variation of this expression with the slip length is small if $L \ll R$.

\end{document}